\documentstyle[prd,epsfig,floats,aps]{revtex}
\begin{document}

\draft

\twocolumn[\hsize\textwidth\columnwidth\hsize\csname @twocolumnfalse\endcsname

\title{Numerical evolution of Brill waves}

\author{David Garfinkle\cite{dgemail}} 
\address{
\centerline{Department of Physics, Oakland University,
Rochester, Michigan 48309}}

\author{G. Comer Duncan\cite{gcdemail}}
\address{
\centerline{Department of Physics and Astronomy, Bowling Green State University,
Bowling Green, Ohio 43403}}

\date{\today}

\maketitle

\begin{abstract}

We report a numerical evolution of axisymmetric Brill waves.  
The numerical algorithm
has new features, including (i) a method for keeping the metric 
regular 
on the axis
and (ii) the use of coordinates that bring spatial infinity to the edge of the
computational grid.  The dependence of the evolved metric on both the amplitude 
and shape of the initial data is found.
 
\end{abstract}


\narrowtext

\vskip2pc]

\section{Introduction}

In the last decade, there has been much use of numerical simulations to study
gravitational collapse.  These studies include critical 
collapse\cite{matt,carsten}, black
hole collisions\cite{ncsa1,ncsa2,ncsa3} and the approach to the 
singularity\cite{beverly1,beverly2,bevprl,allofus}.  In simulations,
spherical or planar symmetry allows the finest resolution, while complete
lack of symmetry (a full 3+1 simulation) 
would allow one to treat a completely
general situation.  An intermediate case is axisymmetry, which allows
more resolution than a full 3+1 simulation; but also permits the study of
situations ({\it e.g.} prolate collapse or black hole collisions) that do
not occur in spherical symmetry.  

One issue that can be addressed in axisymmetry is weak cosmic censorship:
whether the singularities formed in gravitational collapse are hidden inside
black hole event horizons.  A simulation done by Shapiro and Teukolsky\cite{st}
of the collapse of collisionless matter indicates that weak cosmic censorship
may be violated in the collapse of highly prolate objects.  The result of 
reference\cite{st} is that for highly prolate initial configurations, a 
singularity forms before an apparent horizon.  This result might still be
consistent with weak cosmic censorship if it is 
an artifact of the slicing\cite{wald}
or the type of matter used.  To address the second possibility,  
one would like to know whether the same type of behavior occurs 
in vacuum collapse.

In \cite{ast} Abrahams {\it et al} examine families of initial data
for Brill waves.\cite{Brill}  These are vacuum, axisymmetric initial
data at a moment of time symmetry.  The authors of \cite{ast} show that by
considering sufficiently prolate configurations, one can find initial
data with no apparent horizon but with large values of the Riemann
invariant ${R^{abcd}}{R_{abcd}}$.  They conjecture that such initial
data, when evolved, will form singularities without apparent horizons.

In order to test this conjecture, one would like to evolve initial
data for highly prolate Brill waves to find the behavior of the
evolved spacetime.  More generally, one would like to know how the
collapse process depends on both the shape and the strength of the
initial data.
In this paper, we report numerical simulations of the collapse of Brill
waves.  We find the dependence of
the collapse on the amplitude and the initial shape of the wave.  
The numerical 
method is presented in section 2 and the results in section 3.
Conclusions are given in section 4.

\section{Numerical method}

One difficulty present in numerical simulations of axisymmetric spacetimes
is the existence of the axis where the Killing field vanishes.  If one chooses
coordinates adapted to the symmetry, then there is a coordinate singularity
on the axis.  This coordinate singularity has the possibility of causing
numerical problems: that is, one must be very careful in the choice of
numerical method to ensure that the metric remains smooth on the axis.
Most axisymmetric simulations use spherical coordinates.  In addition to 
a coordinate singularity 
on the axis, spherical coordinates lead to a more
severe coordinate singularity at the origin.  This singularity is either 
avoided by using initial data with a black hole and no origin,\cite{ncsa1}
or treated by using elaborate numerical methods to keep the 
metric regular at the
origin.\cite{evans}

We bypass this origin difficulty by the use of cylindrical 
coordinates $(z,r,\phi )$.
The spatial metric ${\gamma _{ab}}$ takes the form
\begin{equation}
\label{gamma}
d {s^2} = {\psi ^4} \left [ {e^{2 r S}} \left ( d {z^2} + d {r^2}
\right ) +  {r^2} d {\phi ^2} \right ]
\end{equation}
Here, $\psi $ and $S$ are functions of $z, \, r$ and the time $t$.  The axis is
at $r=0$ and is the only coordinate singularity.

Numerically, the axis is an edge of the computational grid, 
and therefore values
for the variables must be given 
on the axis points.  However, analytically
the axis consists of interior points, and therefore the only permissable
condition to impose is smoothness.  For a scalar $f$, smoothness requires
that $f$ be an even function of $r$ and therefore that ${\partial _r} f$ vanish
on the axis.  This condition can be used numerically to set the value of $f$ at
the axis points.  Similarly, for components of tensors 
in cylindrical coordinates,
smoothness requires either that the component be an even function of $r$ or
that it be an odd function of $r$.  Odd functions of $r$ 
vanish 
on the axis, while
even functions have vanishing $r$ derivative 
on the axis.  In either case this
information can be used to set the value of the variable 
at the axis points.

Unfortunately, a difficulty arises because smoothness requires that certain 
quantities be even and also vanish 
on the axis.  One such quantity is 
$ {{K^r}_r} - {{K^\phi}_\phi}$.  This quantity must both vanish and have
vanishing derivative 
on the axis.  However, since we can impose only one
boundary condition at the edge of the computational grid, we do not have a
way to maintain both conditions.  Our solution to this difficulty is to 
introduce a new variable: define $W \equiv ( {{K^r}_r} - {{K^\phi}_\phi})/r$.
Then $W$ is odd and smoothness requires only that $W$ vanish 
on the axis.
This method is the reason that the term in the exponential in 
equation (\ref{gamma}) is 
written as $2 r S$.  Smoothness of the spatial metric requires that 
$ {\gamma _{\phi \phi}}/({r^2} {\gamma _{rr}}) = 1 + o({r^2})$ and therefore 
that the quantity $rS$ both vanish and have vanishing derivative 
on the axis.
This condition is met if $S$ vanishes 
on the axis.  In the end, all quantities
we deal with either vanish 
on the axis or have vanishing derivative there, 
{\it but not both}.  If we encounter a quantity $f$ that is order $r^2$ at the
axis, we simply define the odd quantity $g=f/r$ and rewrite all equations 
containing $f$ in terms of $g$.

There remains the question of how to numerically implement the appropriate 
conditions 
on the axis points.  For an odd quantity $X$, the most natural
method would be to put the first grid point at $r=0$ and impose the 
condition $X(1)=0$.  However, for an even quantity $Y$, the simple condition
$ Y(1)=Y(2)$ would then impose ${\partial _r} Y = 0 $ not 
on the axis but
at $r={\Delta _r}/2$ where $\Delta _r$ is the grid spacing in $r$.  Instead
we use the method of ``ghost zones'': the first grid point 
is at $r=-{\Delta _r}/2$.
The axis is then halfway between gridpoints 1 and 2.  For an odd quantity $X$, 
we impose the condition $X(1)=-X(2)$ while for an even quantity $Y$, we use
$Y(1)=Y(2)$.  In each case, the appropriate condition is satisfied 
on the axis.

We note that our method is not the only way to keep the axis stable.
The ``cartoon'' method of reference\cite{cartoon} begins with a cartesian
3+1 code and operates it in a thin slab with boundary conditions at the
faces of the slab given by the axisymmetry of the solution.  This cartoon
method is effective (see reference\cite{shibata} for another 
implementation).  However, since a 3+1 code uses more variables than
an axisymmetric code, for a given axisymmetric problem the cartoon method
uses more computer memory than our method.

We now turn to the method of evolution.  The spatial metric $\gamma _{ab}$ is
evolved using the ADM equation
\begin{equation}
\label{gdot}
{\partial _t} {\gamma _{ab}} = - 2 \alpha {K_{ab}} + {{\cal L}_\beta }
{\gamma _{ab}}
\end{equation}
where $K_{ab}$ is the extrinsic curvature, $\alpha$ is the lapse and $\beta ^a$ 
is the shift.  From the form of the metric in 
equation (\ref{gamma}) it is clear that we
have imposed the conditions ${\gamma _{rz}} = 0 $ and
${\gamma _{rr}} = {\gamma _{zz}}$.  In order that these conditions be preserved
by the evolution in equation (\ref{gdot}), the shift must satisfy
\begin{equation}
{\partial _r} {\beta ^z} + {\partial _z} {\beta ^r} = 2 \alpha
{{K^z}_r} 
\end{equation}
\begin{equation}
{\partial _z} {\beta ^z} - {\partial _r} {\beta ^r} = \alpha U 
\end{equation}
where $ U \equiv {{K^z}_z} - {{K^r}_r}$.   This gives rise to the equations
\begin{equation}
\label{betar}
{\partial _r} {\partial _r} {\beta ^r} + {\partial _z} {\partial _z} {\beta ^r} 
= 2 {\partial _z} (\alpha {{K^z}_r}) - {\partial _r} ( \alpha U ) 
\end{equation}
\begin{equation}
\label{betaz}
{\partial _r} {\partial _r} {\beta ^z} + {\partial _z} {\partial _z} {\beta ^z} 
= 2 {\partial _r} (\alpha {{K^z}_r}) + {\partial _z} ( \alpha U ) 
\end{equation}

Equation (\ref{gdot}) yields the following evolution equation for $S$
\begin{equation}
\label{evolveS}
{\partial _t} S = - \alpha W + {\beta ^z} {\partial _z} S
+ {\beta ^r} {\partial _r} S + {\beta ^r} S/r +
{\partial _r}  ( {\beta ^r}/r ) 
\end{equation}
where $W \equiv ({{K^r}_r}-{{K^\phi}_\phi})/r$.
Rather than evolve $\psi $, we solve for it using the Hamiltonian constraint.

We choose maximal slicing ($K=0$) and evolve the extrinsic curvature using the
ADM equation
\begin{equation}
\label{kdot}
{\partial _t} {{K^a}_b} = - {D^a} {D_b} \alpha + \alpha {{R^a}_b}
+ {{\cal L}_\beta } {{K^a}_b}
\end{equation}
where $R_{ab}$ is the Ricci tensor of the spatial metric.  Since $K=0$, the only
independent components of the extrinsic curvature are ${{K^z}_r}, \, U$ and $W$.
These evolve as follows:
\begin{eqnarray}
\nonumber
{\partial _t} {{K^z}_r} = {\psi ^{-4}} {e^{ - 2 r S}} \left [  
\left ( S +  2 {\psi ^{-1}} {\partial _r} \psi  +  
r {\partial _r} S\right ) {\partial _z} \alpha  \right . \\
\nonumber
\left . - {\partial _z} {\partial _r} \alpha 
+ \left ( 2 {\psi ^{-1}} {\partial _z} \psi 
+ r {\partial _z} S \right ) {\partial _r} \alpha + 
\alpha {R_{zr}} \right ] \\ +  {\beta ^z} {\partial _z} {{K^z}_r} +  
{\beta ^r}
{\partial _r} {{K^z}_r} + {1 \over 2} U \left ( {\partial _r} {\beta ^z}  -
{\partial _z} {\beta ^r} \right ) 
\label{evolveKzr}
\end{eqnarray}
\begin{eqnarray}
\nonumber
{\partial _t} U = {\psi ^{-4}}  {e^{ - 2 r S}}   \left [ 
\left (
4 {\psi ^{-1}}  {\partial _z} \psi + 2 r {\partial _z} S
\right ) {\partial _z} \alpha \right . \\
\nonumber
-  \left ( 2 S + 4 {\psi ^{-1}} {\partial _r} \psi 
+ 2 r {\partial _r} S
\right ) {\partial _r} \alpha  \\
\nonumber \left . + {\partial _r} {\partial _r} \alpha
- {\partial _z} {\partial _z} \alpha - \alpha {R_a} \right ] \\ 
+ {\beta ^z} {\partial _z} U + {\beta ^r} {\partial _r} U 
+ 2 {{K^z}_r}  \left ( {\partial _z} {\beta ^r}  -  {\partial _r}
{\beta ^z} \right )
\label{evolveU}
\end{eqnarray}
\begin{eqnarray}
\nonumber
{\partial _t} W = {\psi ^{-4}}  {e^{- 2rS}} \bigl [ - {\partial _r}
\left ( {r^{-1}} {\partial _r} \alpha \right ) - {\partial _z} S
{\partial _z} \alpha  \\
\nonumber  + \left ( S / r + {\partial _r} S
+ 4 {{(r\psi )}^{-1}} {\partial _r} \psi  \right ) 
{\partial _r} \alpha \bigr ] \\
\nonumber
+ \alpha {R_b} + {\beta ^z} {\partial _z} W + {\beta ^r}
{\partial _r} W + W {\beta ^r}/ r \\
+ ({{K^z}_r}/ r)
\left ( {\partial _r} {\beta ^z} - {\partial _z} {\beta ^r}
\right ) 
\label{evolveW}
\end{eqnarray}
Here we have $ {R_a} \equiv {R_{rr}} - {R_{zz}} $ and $ {R_b} \equiv
( {{R^r}_r} - {{R^\phi }_\phi} )/r$.

The evolution must preserve the condition $K=0$, which implies that the lapse
must satisfy ${D_a} {D^a} \alpha = \alpha {{K^a}_b}{{K^b}_a}$.  This equation
becomes
\begin{eqnarray}
\nonumber
{r^{-1}} {\partial _r}
\left ( r {\psi ^2} {\partial _r} \alpha \right ) + {\partial _z}
\left ( {\psi ^2} \; {\partial _z} \alpha \right ) 
\\ = \alpha {\psi ^6} {e^{2 r S}}  \left [ {2 \over 3}  ( {U^2}
+ {r^2} {W^2} + U r W ) \, + \, 2 {{({{K^z}_r})}^2}\right ]
\label{lapse}
\end{eqnarray}

The Hamiltonian constraint, $R - {{K^a}_b}{{K^b}_a}=0$ becomes the following
equation for the conformal factor $\psi$.
\begin{eqnarray}
\label{hamcon}
\nonumber
({\partial _r} {\partial _r}  + {r^{-1}} {\partial _r}  + {\partial _z}
&{\partial _z}& ) \psi  \\ 
\nonumber = - (\psi /4) \biggl [
({\partial _r} {\partial _r} + {\partial _z} &{\partial _z}& 
) (r S) \\ 
+ {\psi ^4} {e^{2 r S}} \biggl ( {1 \over 3}  ( {U^2} 
+ {r^2} {W^2} + U r &W& )  +   {{({{K^z}_r})}^2}\biggr ) \biggr ]
\end{eqnarray}

Our set of variables is then 
($S,\psi,{{K^z}_r},U,W,\alpha,{\beta^r},{\beta^z}$).
Of these variables, $S, {{K^z}_r}, W$ and $\beta ^r$ 
are odd functions of $r$, 
while the rest are even functions.  These variables are evolved 
as follows: $S$ and the
extrinsic curvature  variables are evolved using equations
(\ref{evolveS},\ref{evolveKzr},\ref{evolveU},\ref{evolveW}). 
At each time step, the elliptic equations 
(\ref{betar},\ref{betaz},\ref{lapse},\ref{hamcon}) 
are solved for the shift, lapse and conformal factor.
  
To begin the evolution, we need initial data satisfying 
the constraint equations.
Initial data for Brill waves is a moment of time symmetry, so ${K_{ab}}=0$ and
therefore the momentum constraint is automatically satisfied.  
The variable $S$ can be freely specified (subject to smoothness 
on the axis
and asymptotic flatness at infinity).  Our choice for $S$ is
\begin{equation}
S = a r \exp \left [ -  {{r^2} \over {\sigma _r ^2}} 
- {{z^2} \over {\sigma _z^2}} \right ]
\end{equation}
where $a, \, {\sigma _r}$ and $\sigma _z$ are constants.  Here, $a$ is the 
amplitude of the wave and $\sigma _r$ and $\sigma _z$ are widths in the $r$
and $z$ directions respectively.  Given $S$, equation (\ref{hamcon}) is solved
for $\psi$. 

There remains the question of the boundary conditions to apply at the outer edge
of the computational grid.  A natural method would be to put the outer edge of
the computational grid at some large distance and to impose some outgoing wave
condition on the evolution equations and a Robin boundary condition on the
elliptic equations.  However, the issue of appropriate boundary conditions for
a mixed hyperbolic-elliptic set of equations is quite complicated.  This issue
becomes even more dificult in cylindrical coordinates than in spherical 
coordinates since the asymptotic behavior of the variables looks more 
complicated in cylindrical coordinates.  Our attempts to impose a boundary
condition of this sort led to numerical instability.  Instead, we decided
to use a different approach.  We begin by noting that a coordinate transformation
can bring spatial infinity to the edge of the computational grid.  We introduce
new coordinates $({\tilde z},{\tilde r})$ defined by $z = \tan {\tilde z}$ and
$r = \tan {\tilde r}$.  We then place the edges of the computational grid at
${\tilde z} = \pi /2$ and at ${\tilde r} = \pi /2$.  These regions correspond
to spatial infinity.  Though we use new coordinates, we retain the old
metric and extrinsic curvature variables, with the exception that we introduce
the quantities ${\tilde S} = S/\cos {\tilde r}$ and ${\tilde W} = W/\cos {\tilde r}$.
Thus our set of variables is 
$({\tilde S},\psi,{{K^z}_r},U,{\tilde W},\alpha,{\beta^r},{\beta^z})$ and the
set of equations used to evolve these variables is equations 
(\ref{evolveS},\ref{evolveKzr},\ref{evolveU},\ref{evolveW},\ref{betar},
\ref{betaz},\ref{lapse},\ref{hamcon}) with the
substitutions $ r \to \tan {\tilde r} , {\partial _r} \to {\cos ^2} {\tilde r}
{\partial _{\tilde r}}, \, {\partial _z} \to {\cos ^2} {\tilde z} {\partial _{\tilde z}},
\, S \to \cos {\tilde r} {\tilde S}$ and $W \to \cos {\tilde r} {\tilde W}$.

Since the edge of the grid is at spatial infinity, the choice 
of boundary condition
is dictated by asymptotic flatness: $\psi $ and $\alpha $ must be 1 at the outer
boundary, and all the other variables must vanish there.  Though the coordinate
transformation is singular, this does not lead to a singularity in the evolution
equations.  In fact, just the opposite is true: in the new coordinates, the
right hand sides of the evolution equations approach zero at spatial infinity
(as they must, since the variables are unchanging there).

The advantages of this spatial infinity boundary condition are stability and
consistency with the field equations.  However, there are disadvantages as well.
The change of variables changes the form of the elliptic operators that need to
be inverted to solve the elliptic equations.  This slows down the process of
solving the elliptic equations and therefore slows down the code.  A more 
fundamental difficulty has to do with waves produced in the collapse process.
As a wave travels outward, its (approximately constant) physical width 
corresponds to fewer coordinate grid spacings.  Eventually, the wave fails to
be resolved, and since the system is mixed hyperbolic-elliptic, this failure
of resolution in one part of the grid can affect the entire grid.  
This difficulty
means that for a given spatial resolution, there is only a 
certain amount of time
that the evolution can be run and still give reliable results.  
This places a limit
on the type of problems that can be treated using this method.

The variables $({\tilde S},{{K^z}_r},U,{\tilde W})$
are evolved using a 3 step iterative Crank-Nicholson (ICN) algorithm.  At each
step of the ICN process, the elliptic equations for
$(\psi,\alpha,{\beta^r},{\beta^z})$
are solved using the conjugate gradient method with Neumann preconditioning.

Though we solve for $\psi$ using the Hamiltonian constraint, we note
that $\psi$ also satisfies an evolution equation.  From equation
(\ref{gdot}) we have
\begin{equation}
\label{psi2}
{\partial _t} \psi = {\beta ^z} {\partial _z} \psi + {\beta ^r}
{\partial _r} \psi + \psi \left [ {{\beta ^r} \over {2 r}} + 
{\alpha \over 6} (U+2 r W)\right ]
\end{equation}
We use equation (\ref{psi2}) as a check on the reliability of the code.
Specifically, we compute a second value of $\psi$ by starting with the
initial data and then evolving using equation (\ref{psi2}).  We then
check that the two values of $\psi$ agree well at $r=z=0$.  We run the
code only as long as this good agreement persists.

The convergence of the code is tested using the momemtum constraint
${p_a}={D^b}{K_{ab}}$ (Since $K=0$).  The Hamiltonian constraint is
solved for $\psi$ and therefore cannot be used as an independent test.
In general, the methods we use would lead one to expect second order
convergence.  However, the boundary conditions at infinity for the 
elliptic equations may be only first order accurate.  Figure 1 shows
the ${\rm L}^2$ norm of $p_z$ plotted as a function of time.  For these 
runs, $a=4$ and ${\sigma_r}={\sigma_z}=1$.  The dashed line corresponds
to 42 gridpoints in the $r$ direction and 42 gridpoints in the $z$ 
direction, while the solid line corresponds to a run with twice the
resolution.  The results indicate second order convergence.

\begin{figure}[bth]
\begin{center}
\makebox[3.0in]{\psfig{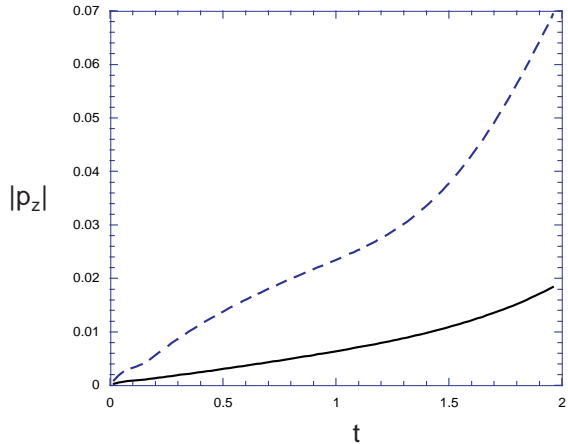}}
\caption{$|{p_z}|$ plotted {\it vs.} $t$ for two 
different resolutions}
\label{fig1}
\end{center}
\end{figure}

\section{Results}

The code was run in double precision on Dec alpha workstations and on
the NCSA Origin 2000.  Unless otherwise stated, all simulations were
run with 42 gridpoints in the $r$ direction and 42 gridpoints in the
$z$ direction.

The initial data that we use, and therefore the spacetime that it evolves
to, has reflection symmetry about the $z=0$ plane.  
Though our algorithm does not require this extra symmetry, since the
spacetimes we treat have this reflection symmetry, we save computational
time by running the  
simulations in the range $0 \le {\tilde z} < \pi /2$.  The reflection 
symmetry and axisymmetry give spacetime a preferred world line: the origin
($r=z=0$).  In gravitational collapse in maximal slicing, it is expected
that the lapse approaches zero in the region of strong gravity.  Therefore,
a useful quantity to plot is $\ln {\alpha _0}$ as a function of time, where
$\alpha _0$ is the lapse at the origin.  Note that this quantity can be
directly compared between two different codes, provided that both use 
maximal slicing.

Our initial data has three parameters: the amplitude $a$ and the two widths
$\sigma _r$ and $\sigma _z$.  However, it turns out that the effective
parameter space is two dimensional.  Under the transformation ${\sigma _r} 
\to c {\sigma _r} , \; {\sigma _z} \to c {\sigma _z} , \; a \to a/{c^2}$ 
for a positive constant $c$, the spacetime is changed only by an overall
constant scale.  We can think of the two remaining parameters as being
the strength and shape of the wave, and we want to find the dependence of
the collapse on these two parameters.  Figure 2 shows the dependence of
the collapse on the strength of the wave.  Here, three simulations are 
run, each with ${\sigma _r} = {\sigma _z} = 1$.  The parameter $a$ is
$4$ for the solid line, $5$ for the dashed line and $6$ for the 
dot-dashed line.  
The $a=4$ and $a=6$ spacetimes have been studied using
a 3+1 code in reference\cite{ed}.  Our results are in good agreement with
theirs.  For $a=4$, the lapse, after an initial collapse to small values,
appears to be returning to 1.  That is, we expect the Brill wave to disperse.
For $a=6$, the lapse continues to collapse, and we expect a black hole to
form.  Somewhere between $a=4$ and $a=6$ is an amplitude that leads to 
critical collapse; however our method does not have the resolution 
needed to study this process.  The reason for this is that to study
critical collapse, one must evolve long enough to distinguish a 
spacetime slightly below the black hole threshold from one that is slightly
above.  In this amount of time, the waves traveling towards the outer
boundary will, in general fail to be resolved, leading to a general 
lack of accuracy in the results of the evolution.   

\begin{figure}[bth]
\begin{center}
\makebox[3.0in]{\psfig{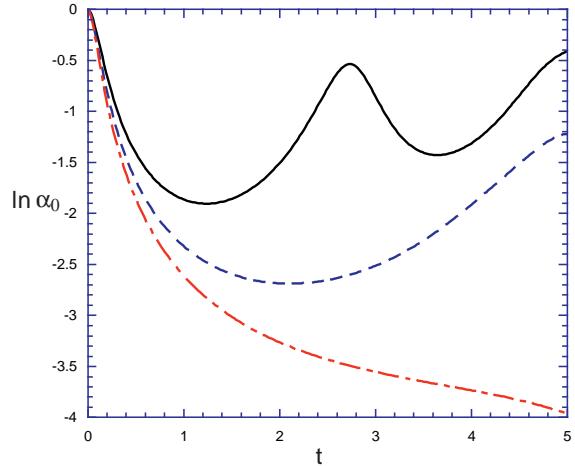}}
\caption{$\ln {\alpha_0}$ plotted {\it vs.} $t$ for three
different amplitudes: $a=4$ (solid line), $a=5$ (dashed line)
and $a=6$ (dot-dashed line)}
\label{fig2}
\end{center}
\end{figure}

The issue of the dependence of the collapse process on the shape is
somewhat subtle, due to the question of what is to be held constant
as the shape varies.  The simplest thing to do is to hold the
amplitude $a$ and the sum ${\sigma _r} + {\sigma _z}$ constant 
while varying ${\sigma _r} - {\sigma _z}$.  Results of these
collapse simulations are shown in figure 3.  Here, $a=4$ and 
${\sigma _r}+{\sigma _z} =2$ for all simulations.  
We will call initial data ``spherical'' if
${\sigma _r} = {\sigma _z}$, ``prolate'' if ${\sigma _r} < {\sigma _z}$
and ``oblate'' if ${\sigma _r} > {\sigma _z}$.  
(This terminology is
somewhat misleading, since it is not the wave itself but the initial
$S/r$ that is spherical, prolate or oblate).  Figure 3 contains a
spherical collapse (solid line), a prolate collapse (dashed line) with
${\sigma _r} - {\sigma _z} = - 0.4$ and an oblate collapse (dot-dashed
line) with ${\sigma _r} - {\sigma _z} = 0.4$.  Here the shape seems
to have a large influence on the collapse, with even a small amount of 
oblateness producing collapse, while a small amount of prolateness 
hastens dispersion.

\begin{figure}[bth]
\begin{center}
\makebox[3.0in]
{\psfig{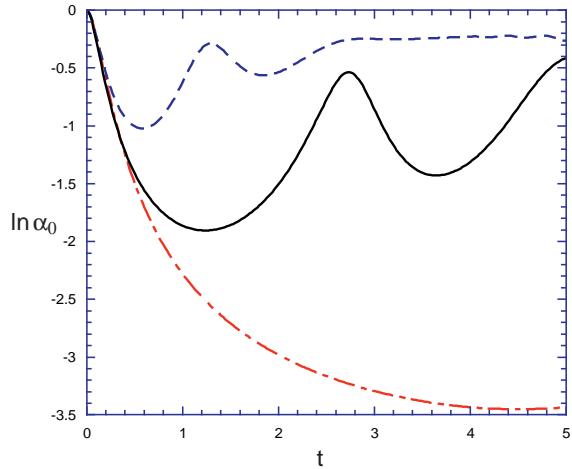}}
\caption{$\ln {\alpha _0}$ plotted {\it vs.} $t$ for $a=4$ and 
three different shapes: spherical (solid line), prolate (dashed
line) and oblate (dot-dashed line)}
\label{fig3}
\end{center}
\end{figure}

However, it is not clear how to interpret this result, since for a fixed
$a$ and ${\sigma _r}+{\sigma _z}$, it may be that there is a dependence of
the ``strength'' of the gravitational wave on ${\sigma _r}-{\sigma _z}$.
To see this, we examine how the ADM mass depends on the degree of prolateness
or oblateness at fixed $a$ and ${\sigma _r}+{\sigma _z}$.  These results
are shown in figure 4.  Here, we have ${\sigma _r}+{\sigma _z}=2$ and
$a=4$.  The ADM mass $M$ is plotted as a function of 
${\sigma _r}/{\sigma _z}$.  Since, at a fixed $a$, $M$ is an increasing
function of ${\sigma _r}/{\sigma _z}$, this is (at least part of) the 
reason why at fixed $a$ oblateness causes collapse.

\begin{figure}[bth]
\begin{center}
\makebox[3.0in]{\psfig{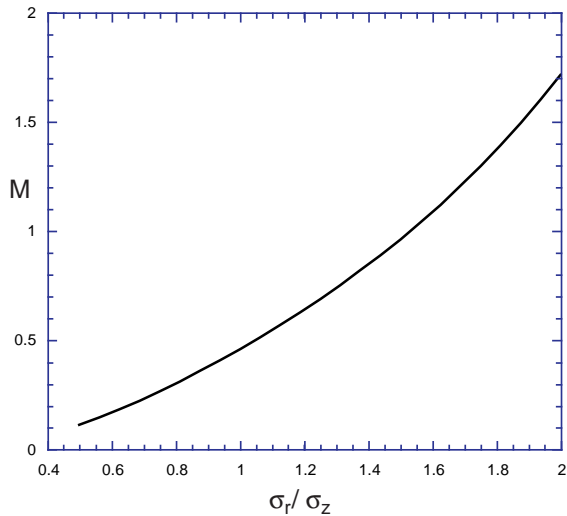}}
\caption{$M$ plotted {\it vs.} ${\sigma_r}/{\sigma_z}$ for 
$a=4$ and ${\sigma_r}+{\sigma_z}=2$}
\label{fig4}
\end{center}
\end{figure}

One might therefore instead look at the influence of shape at fixed 
${\sigma _r}+{\sigma _z}$ and fixed $M$.  The results of this 
comparison are shown in figure 5.  Here ${\sigma _r}+{\sigma _z}=2$
and $M = 0.46 $  (the value of $M$ corresponding to $a=4$ and 
${\sigma _r}={\sigma _z}=1$).  The solid line is a spherical collapse,
while the dashed line is a prolate collapse with ${\sigma _r}-{\sigma _z}
= -0.4$ (corresponding to $a=6.1$) 
and the dot-dashed line is an oblate collapse with 
${\sigma _r}-{\sigma _z}=0.4$ (corresponding to $a=2.7$).  
Here the shape has some influence on
the details of the collapse process; but this much change in the shape
does not seem to have a particular tendency either to promote or to 
inhibit collapse.  Figure 6 shows the same sort of plot, but with 
more distortion in the shape.  Again $M=0.46$ and ${\sigma_r}+{\sigma_z}=2$.
However, here the prolate collapse (dashed line) has 
${\sigma _r}-{\sigma _z}=-1$ ($a=14$) 
and the oblate collapse (dot-dashed line) 
has ${\sigma _r}-{\sigma _z}=1$ ($a=1.4$) 
in addition to the spherical collapse
(solid line). Here it seems that at constant ADM mass there is a slight
tendency of oblateness to induce dispersion. 

\begin{figure}[bth]
\begin{center}
\makebox[3.0in]{\psfig{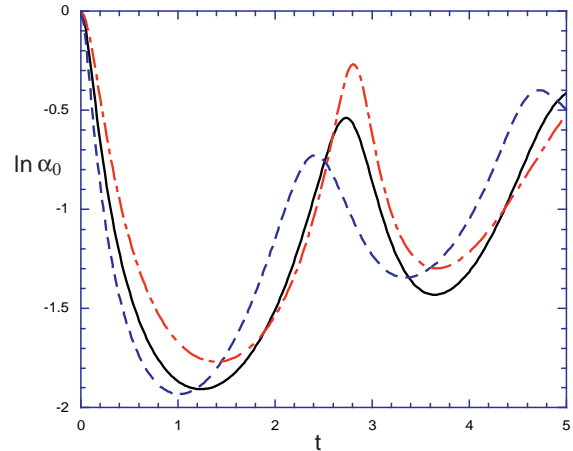}}
\caption{$\ln {\alpha_0}$ plotted {\it vs.} $t$ for $M=0.46$
and three different shapes: spherical (solid line), prolate
(dashed line) and oblate (dot-dashed line)}
\label{fig5}
\end{center}
\end{figure}

\begin{figure}[bth]
\begin{center}
\makebox[3.0in]{\psfig{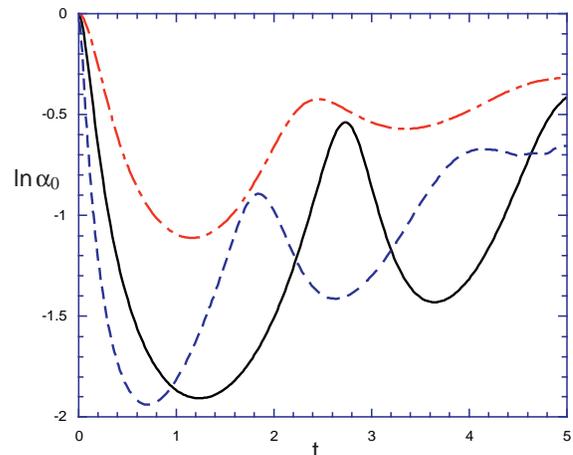}}
\caption{$\ln {\alpha _0}$ plotted {\it vs.} $t$ for $M=0.46$ and
three different shapes: spherical (solid line), prolate
(dashed line) and oblate (dot-dashed line).  Here the amount
of oblateness or prolateness is greater than that in figure 4}
\label{fig6}
\end{center}
\end{figure}

We now consider more strongly gravitating initial data and follow its
evolution until the formation of an apparent horizon, while considering
some of the properties of the collapse.  This simulation has 
$M=2, \, {\sigma _r} =1$ and ${\sigma _z} =1 $ ($a=8.5$) and is run with
82 gridpoints in the $r$ direction and 82 gridpoints in the
$z$ direction.  Figure 7 shows
$\ln {\alpha _0}$ as a function of time.  The lapse collapses
throughout the evolution.  At each time step we calculate the
Riemann invariant $I = {R^{abcd}}{R_{abcd}}/16$ and find the maximum
of its absolute value $I_{\rm max}$ as well as the spatial position
where $|I|={I_{\rm max}}$.  Figure 8 shows $\ln {I_{\rm max}}$ 
plotted as a function of time.  Here we see that after an initial
increase, $\ln {I_{\rm max}}$ decreases during the rest of the
evolution.   
At all times during the evolution the place where $|I|={I_{\rm max}}$
is the origin.  

To monitor
the approach to apparent horizon formation, we consider how the 
horizon is found.  On a maximal slice in an axisymmetric spacetime,
the horizon is given by a curve $z=f(r)$ where the function $f$ 
satisfies a differential equation that is integrated from the axis
to $z=0$.  Let $\theta $ denote the angle at which the curve meets
the $z=0$ plane.  Then the curve is a horizon only if $\theta = \pi /2$.
The horizon finding subroutine integrates the differential equation 
for $f$ starting at each point on the axis, finds the angle $\theta $
for each curve and then finds $\theta _{\rm max}$, the maximum value
over all curves of this angle.  If ${\theta _{\rm max}}< \pi/2$ then
there is no horizon at this time.  Figure 9 shows $ 2 {\theta _{\rm max}}
/\pi$ plotted as a function of time.  Note that $\theta _{\rm max}$ 
increases throughout the collapse process.  Note also that even before
the actual horizon formation, one can tell that a horizon is about to
form by noticing that $\theta _{\rm max}$ is approaching $\pi/2$.   
The horizon forms at $t=3.9$ with area $A=165=0.82(16\pi {M^2})$.

\begin{figure}[bth]
\begin{center}
\makebox[3.0in]{\psfig{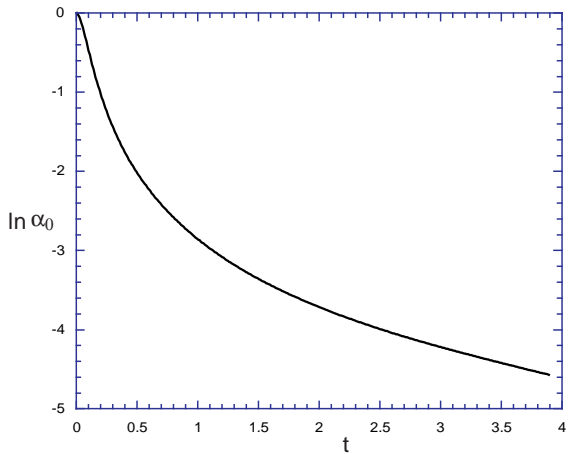}}
\caption{$\ln {\alpha_0}$ plotted {\it vs.} $t$ for
$M=2$ and ${\sigma_r}={\sigma_z}=1$}
\label{fig7}
\end{center}
\end{figure}

\begin{figure}[bth]
\begin{center}
\makebox[3.0in]{\psfig{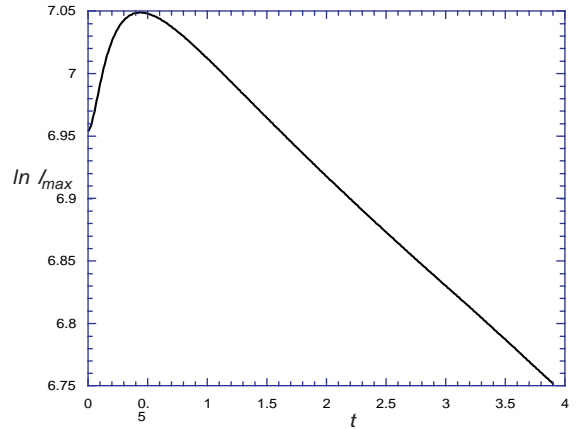}}
\caption{$\ln {I_{\rm max}}$ plotted {\it vs.} $t$ for
$M=2$ and ${\sigma_r}={\sigma_z}=1$}
\label{fig8}
\end{center}
\end{figure}

\begin{figure}[bth]
\begin{center}
\makebox[3.0in]{\psfig{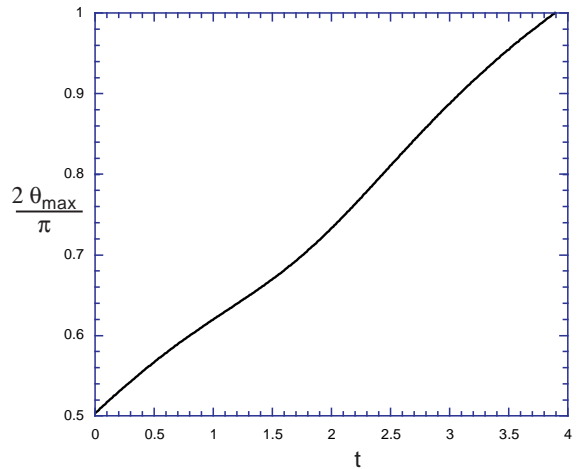}}
\caption{$2 {\theta _{\rm max}}/\pi$ plotted {\it vs.} $t$ for $M=2$
and ${\sigma_r}={\sigma_z}=1$} 
\label{fig9}
\end{center}
\end{figure}

\begin{figure}[bth]
\begin{center}
\makebox[3.0in]{\psfig{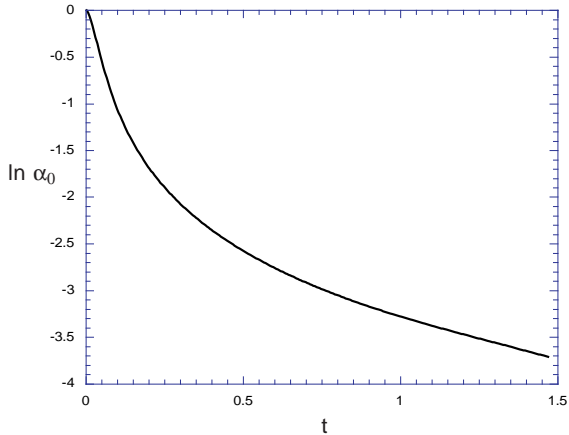}}
\caption{$\ln {\alpha _0}$ plotted {\it vs.} $t$ for
$M=2, \, {\sigma_r}=0.128$ and ${\sigma_z}=1.6$}
\label{fig10}
\end{center}
\end{figure}

\begin{figure}[bth]
\begin{center}
\makebox[3.0in]{\psfig{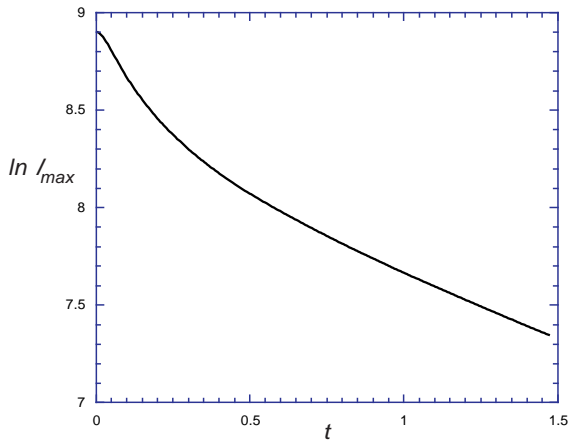}}
\caption{$\ln {I_{\rm max}}$ plotted {\it vs.} $t$ for
$M=2, \, {\sigma_r}=0.128$ and ${\sigma_z}=1.6$}
\label{fig11}
\end{center}
\end{figure}
We now turn to a highly prolate collapse: the evolution of one
of the initial data sets of reference\cite{ast}.  Here, $M=2, \,
{\sigma _z}=1.6$ and ${\sigma _r}=0.128$ ($a=325$).  
This simulation was run
with 162 grid points in the $r$ direction and 42 grid points in the
$z$ direction.  Ideally, we would like to follow the collapse until
either a horizon forms or a curvature scalar blows up.  Unfortunately,
we are not able to follow the evolution that long; so we evolve for
a somewhat shorter time and attempt to discern trends in the evolution.
Figure 10 shows $\ln {\alpha _0}$ plotted as a function of time.  Here
we see the usual collapse of the lapse throughout the evolution.
Figure 11 shows $\ln {I_{\rm max}}$ plotted
as a function of time.  Here we see that the maximum of the 
Riemann invariant decreases as the collapse proceeds.  In the initial
data, the spatial location where $|I|={I_{\rm max}}$ is on the axis
at $z=1.04$.  As the evolution proceeds, this spatial location remains
on the axis, but moves towards the origin, reaching the origin at
$t=0.55$ and then remaining at the origin for the rest of the evolution.
To discern a trend in the approach to apparent horizon formation,
we plot (Figure 12) $2{\theta _{\rm max}}/\pi$ as a function
of time for this simulation.  Note that this quantity is increasing.

The trends of this evolution are that (i) $I_{\rm max}$ decreases, (ii)
the spatial position where $|I|={I_{\rm max}}$ moves to the origin
and (iii) $\theta _{\rm max}$ increases.  If these trends continue, then
this spacetime will form a black hole rather than a naked singularity.
Thus (in this example at least) it seems that Brill waves behave 
differently from collisionless matter with even highly prolate initial
configurations forming black holes.  This conclusion is not firm for 
two reasons: (i) we have only followed the evolution for a certain amount
of time, and the trends that we have observed in this part of the 
evolution could reverse in later parts.  (ii) we have only evolved a
certain, highly prolate initial configuration.  It is possible that
much more prolate initial configurations behave differently.  Both
these issues need further study. 

\begin{figure}[bth]
\begin{center}
\makebox[3.0in]{\psfig{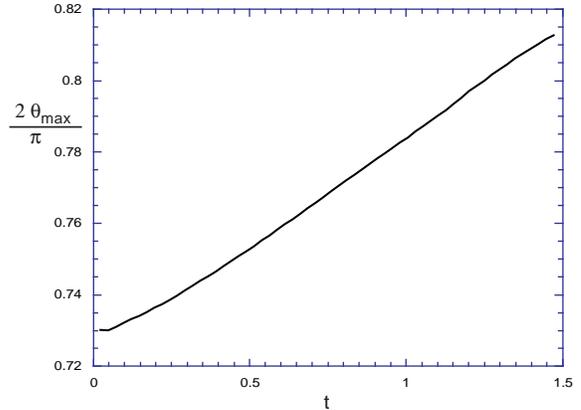}}
\caption{$2 {\theta _{\rm max}}/\pi$ plotted {\it vs.} $t$ for
$M=2, \, {\sigma_r}=0.128$ and ${\sigma_z}=1.6$}
\label{fig12}
\end{center}
\end{figure}

\section{Conclusion}

We have presented (i) a new method of curing axis instabilities, 
(ii) a study of the dependence of Brill wave collapse on the shape
of the wave including (iii) a preliminary examination of the issue
of cosmic censorship in the collapse of highly prolate Brill waves.
A convincing demonstration that the axis really is stable would
require evolution for long times.  However, our outer boundary
condition entails a loss of resolution that increases with time
and this prevents us from running the code for a long time, even
for weak waves.  (Note: one might also expect that our sort of outer
boundary condition would result in spurious reflected waves.  While
we cannot run our code long enough to check this, we have used our
outer boundary condition to perform a numerical simulation of the linear
wave equation; and there we find very little reflection).

A way to test the regular axis method independently of the question of
outer boundary conditions is to treat the case of closed cosmologies.
One of us has used the regular axis method to treat Gowdy spacetimes
on ${S^2}\times {S^1} \times R$.\cite{gowdys2}  Here the evolution
proceeds for long times and there is no axis instability.  Though the
Gowdy spacetimes have two Killing fields, one can easily apply the regular
axis method to closed axisymmetric cosmologies. 

To do a more thorough study of Brill wave collapse, we require a better
way of treating the outer boundary.  One way to do this is to put the
boundary at a finite distance but replace the ADM system by one in
which a simple outgoing wave boundary condition does not cause numerical
instability.  The method of Baumgarte, Shapiro, Shibata and
Nakamura\cite{bssn,sn} seems to have the appropriate properties.
Another method would use harmonic coordinates, which makes Einstein's
equation similar to the wave equation and should be stable with an
outgoing wave boundary condition.  Alternatively, one could keep
spatial infinity (or null infinity) on the computational grid, but
replace the ADM equations with the conformal equations of 
Friedrich.\cite{hf1,hf2,ph}  We are presently examining these alternatives
to see which is likely to work best for Brill wave simulations. 

Our preliminary results on Brill wave collapse indicate that even 
highly prolate Brill waves do not evolve to form naked singularities.
In the evolution of highly prolate initial data, gravitational
collapse will tend to increase the Riemann curvature, while
the tendency of the wave to spread out will lessen curvature.
We have shown that in the initial stage of the evolution of a
highly prolate wave, the second effect is the more important one.
That is, the highly prolate wave tends to become less prolate
rather than more as it evolves.  In this way, Brill waves behave
differently from the collisionless matter studied in
reference\cite{st}.  To put our tentative conclusions on a firmer
footing, we need to study more extreme configurations and evolve
them for a longer time. 

\acknowledgments
We would like to thank Stu Shapiro, Thomas Baumgarte, Peter Huebner, 
Abhay Ashtehar
and especially Matt Choptuik, Beverly Berger and Masaru Shibata for
helpful discussions.
We would also like to thank the authors of reference\cite{ed} for
making their data available to us.
This work was partially supported by NSF grant PHY-9722039 to 
Oakland University.  Some of the simulations were performed at
the National Center for Supercomputing Applications (Illinois).

\end{document}